Multiplayer Games and their Need for Scalable and Secure State Management
Zakaria Alomari
Department of Computer Science & Software Engineering
Concordia University
Montreal, Quebec Canada

# Abstract

In recent years, massively multiplayer online games (MMOGs) have become very popular by providing more entertainment, therefore millions of players now participate may interact with each other in a shared environment, even though these players may be separated by huge geographic distances. Peer to Peer (P2P) architectures become very popular in MMOG recently, due to their distributed and collaborative nature, have low infrastructure costs, achieve fast response times by creating direct connections between players and can achieve high scalability. However, P2P architectures face many challenges and tend to be vulnerable to cheating. Game distribution between peers makes maintaining control of the game becomes more complicated. Therefore, broadcasting all state changes to every player is not a viable solution to maintain a consistent game state in a MMOGs. To successfully overcome the challenge of scale, MMOGs have to employ sophisticated interest management techniques that only send relevant state changes to each player. In this paper, In order to prevent cheaters to gain unfair advantages in P2P-based MMOGs, several cheat-proof schemes have been proposed that utilize a range of techniques such as cryptographic mechanisms, Commitment and agreement protocols, and proxy architecture.

# Table of Contents

# List of Abbreviations

**MMOGs** : Massively Multiplayer Online Games.

**P2P :** Peer-to-peer architectures.

**FPS :** First-person shooter.

**TCGs :** Trading Card Games.

**IRS :** Invigilation Reputation Security.

# 1. Introduction

The popularity of Massively Multiplayer Online Games (MMOGs) has been increased within the last years, now with more than 12 million subscribers [1] and market value revenues exceeding $1 billion annually [2].

Basically MMOGs are a multiplayer video game (computer game) where is capable of supporting large numbers of players simultaneously and interact with each other online in a persistent virtual world [3,4]. On the other hand, players not only interact with each other and the virtual world, but sometimes also participate in building the virtual world itself. A traditional multiplayer game in which usually up to ten of players play a relatively short-lived game, as compared with MMOGs which offer the possibility for tens of thousands of players to play together in a persistent world (long-living) [5]. When one player (client) performs an action such as acquiring and improving skills, picking up items in his/her inventory … , etc. that affects the world, thus, the game state of all other players affected by that action must be updated. Therefore, these games are meant to be played for a long period of time, with users spending several months or years playing on a single character.

MMOGs considered a huge database because there are a lot of states and a lot of actions on the data, it is a distributed architecture by nature which make it interesting passes security aspect. Thus, the main challenges in MMOGs are scalability, consistency and security.

There are two main paradigms of architecture in use for MMOGs: client-server, and peer-to-peer (P2P). In a client-server architecture (see Figure 1) the server can be one or a cluster of dedicated machines that are usually maintained by the game provider. In other words, client-server architecture, a single server provides the entire game environment for all clients or some subset of clients. In a P2P architecture (see Figure 2), peers may be connected with an arbitrary number of other peers. Further, there is no central point of control. The peers in the game are used as resources to run and manage the game in a distributed fashion [6,7]. *P2P architectures have received a great deal of research attention in the recent past as they distribute computational and network load among peers, can potentially achieve high scalability, low cost, and good performance*[10].

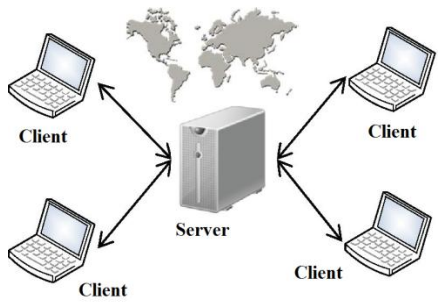 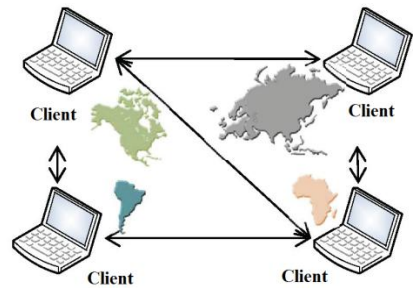

**Figure 1:** Client-Server architecture            **Figure 2**: peer-to-peer architecture

## 1.1 Problem Statement

The development of MMOGs comprise a number of challenges when considering a P2P architectures, as a result of their design, they are more susceptible to attacks and cheating than centralized systems. One of the serious matters with P2P architectures is security. Therefore, cheating is easier in a P2P environment. The reason behind that is a P2P system might elect to place decision making authority on user machines, providing opportunity to cheat [5]. On the other hand, cheating is commonly one of the essential concerns in the design of a game architecture and one of the reasons game companies have avoided the use of P2P architectures.

Furthermore, Most cheating occurs by players to gain some undue advantage over other players and to speed up their progress in the game. This affect the honest players to be frustrated and either becoming a cheater themselves or tend to leave a game in which they perceive themselves to be at a disadvantage due to widespread cheating [9]. In addition, cheating techniques could be established on collaboration where several players collaborate for cheating purposes. For example, *if the P2P overlay is used for update dissemination, players can cheat by withholding or accessing update messages, giving them an unfair advantage. If the P2P architecture provides distributed execution of updates and allows players to hold primary copies of game objects, cheaters can manipulate their object repositories and perform game state changes that violate game rules* [10].In other words, Cheating occurs when a player makes changes to State Game defying the rules and lead to an unfair advantage.

## 1.2 Motivation

The main motivation of our study is trying to understand how MMOGs actually allow large groups of people to play a single online game simultaneously, and what are the problems that are still faced the players while playing this game. Furthermore, there was another motivation, which is the architectural nature of cheating, given our interest in the impact of P2P on the vulnerability of multiplayer games. While many of these cheating techniques can occur even in client-server systems, but cheating is easier in a P2P environment. However, other motivations will be introduced as such as the techniques which used to prevent the happening of cheating, also detection techniques which means reactive and rely on the fact that if cheaters know that they will get caught and be punished.

## 1.3 Objective

Following the problems mentioned above, our main objective is to design a scalable, cheat-resistant, and fast architecture. The reason of focusing on the cheat prevention and proactive approaches because we want prevention techniques proactively reduce or eliminate the possibility of cheating by players.

# 2. Background

Cheating, is an activity that changes the game experience to give advantage to one player than others. Although cheating has been reported in the multiplayer online games, it is difficult to measure. Cheating technology, can be classified along several dimensions or different categories of cheating. The first category is Interrupting Information Dissemination, which works through changing the update rate or sending an incorrect or inaccurate information. Therefore a player will be able to confuse the other players in the current state. As a result, the chance of his/her avatar to win the unfair advantage in his/her attacks. More specifically, the nodes in the P2P architecture is more vulnerable to this type of cheating, the reason behind that they are in charge of disseminating their own updates as well as those of others. In other words, these nodes that are part of a forwarding pool , will be able to interrupt information dissemination easily. There are various types of cheating are classified in this category such as : Escaping, Time cheating, Network flooding, Fast rate cheat, Suppress-correct chat, replay cheat and Blind opponent. More detailed, one of these types is Time Cheating which uses the fake timestamps of the past , where a player, after receiving an updates from the others, and then sends his own update with an old timestamp to avoid detection. On the other hand, there is a Replay Cheat, in this kind *a cheater resends signed and encrypted updates of a different player that she has previously received*[10]. Furthermore, a cheater gets updates from the opponents, although blinding them about the cheater's actions; by drops some updates to opponents, this is called Blind Opponent [8,10].

The second category is Illegal Game Actions, which work through circumvent the game physical laws by cheater and unduly change his state by tampering with the game code. There are various types of cheating are classified in this category such as : Client-side code tampering, Aimbots, Spoofing and Consistency cheat. More detailed, An aimbot is a type of computer game bot used in multiplayer first-person shooter (FPS) games to provide varying levels of target acquisition assistance to the player. In this kind, the player, in order to provide his/her automatic weapon aiming, uses intelligent program. On the other hand, the cheater sends different updates to different players , in addition this kind can be used by a player or a group of players this is called Consistency cheat [10].

The last category is Unauthorized Information Access, which works through that players will be able to take advantage of the available information, it should not be disclosed to increase the ability to kill another player, thus helping him evade. There are various types of cheating is classified in this category such as : Sniffing, Maphack and Rate analysis. More detailed , Maphack is a generic term that refers to a method or third-party program that enables a player to see more of a level than intended by the developer. A maphacker is a player who deliberately executes such a method or program in the context of a relevant game, whilst maphacking is the act of such [10,11].

## 3. Technical Aspects

There are a number of different techniques employed to address a limited number of cheat types that may occur in certain architectures and scenarios in MMOGs. Some techniques try to prevent cheating from happening, while others detect cheating. Furthermore, there are some existing engines that use P2P solutions and that we believe can be extended for P2P support.

## 3.1 Cheating Prevention

To prevent cheating, games may espouse cryptographic mechanisms, such as message encryption, signatures, and checksums are effective in eliminating message sniffing and illegal message modifications. Cryptographic techniques have been proposed to prevent players' actions from being known by others before each player submits the final decisions [12]. For example, *cryptographically secure hashes of cards has been used to secure Trading Card Games (TCGs) eliminating the need for a referee (later explained)* [10]. Another mechanism to prevent cheating is Commitment and agreement protocols. In this mechanism, we adopted on a secured event updating protocol which includes: a commitment scheme and a digital signature scheme. After the decision has been made, the commitment scheme ensures that the players will not change their behaviors (actions). On the other hand, a digital signature scheme to ensure that players will not be able to deny the act that they have done. If the commitments are digitally signed, we can also prevent impersonation and dodging [12]. For example, *time cheating is addressed by the lockstep protocol It requires all players to first submit a hash code of their next actions and only after every player's hash code has been received, players send their actions to each other. By comparing the hash code with the action, players can make sure that other players have not changed their actions after receiving input from others*[10]. Another popular method to prevent cheating is by proxy architecture (see Figure 3) where proxies forward messages and manage subscriptions of the interest sets of players. Furthermore, proxies also hide dead-reckoning information from players unless they are in the vision range [10,13].

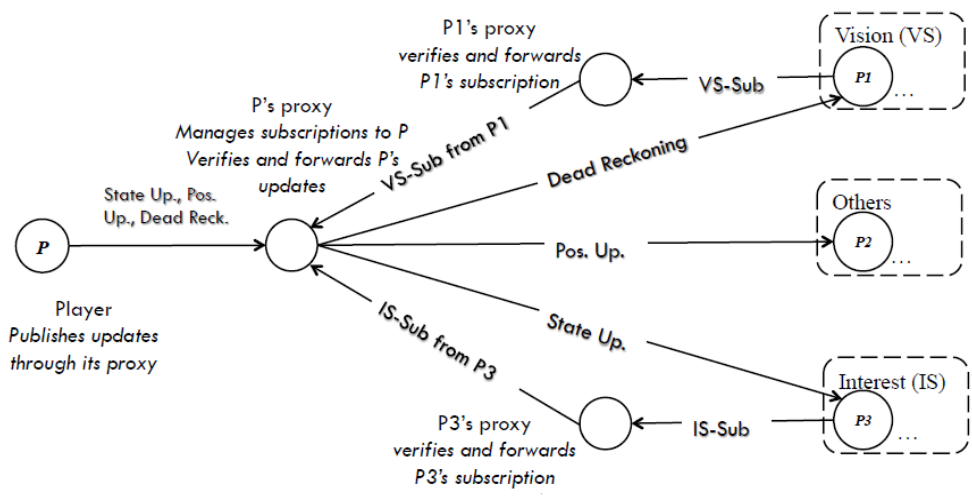

**Figure 3:** Watchmen: Proxy Scheme [13]

## 3.2 Cheating Detection

As mentioned previously, cheating represents a serious threat. Therefore, this problem is exacerbated if no adequate methods are found to be detected and sanction cheating players. Sanction typically constitutes being banned from playing for a certain duration or indefinitely. Detection techniques dealing with cheating in P2P games can be classified such as Game Log Verification which is an analysis technique to find anomaly or detect bots, in other words, *a common technique is that all actions are audited and verified for security breaches* [10]. One of the forms of verification that can detect cheaters is comparing hash messages of future updates with the actual updates. For example, at the start of the game, each player, it is necessary to replace the hash value of the initial state of the game, at the end of the game, each player exchanges all the operations he issued and then to confirm the validity of the hash value and the state / operations to simulate the game again [8,10].

Another mechanism is a Referee Selection. Therefore, in the coordination-based architecture, there is a centralized authority in each region, that is, the super-peer; some peers are elected to be super-peers, which operate both as a server to a set of clients, and as an equal in a network of super-peers. on the other hand, this node can perform a most of the security checks, assuming it to have a high reliability [14]. Another hybrid approach is Invigilation Reputation Security (IRS) to purposes of controlling and eliminating game cheaters. Where communication is handled by the server and update execution is managed by peers. In this approach the server assigns a proxy peer to each peer. Proxy peers are selected at random, and will be assigned on a regular basis. All the executed updates and returning results of the peers are done by the proxy peer, on the other hand, the sever is responsible to relay the message between the peer and its proxy. Server runs a quick revision test to see If the result is returned while if it is possible according to the rules or not. Conflicting updates and a percentage of updates randomly chosen are then re executed by a selected monitor peer to verify the results and detect cheating [10]. Moreover, Mobile Guards are used to ensure the integrity of the protection mechanism, which aims at preventing cheating through modification of game client. Mobile guard is a small code segments downloaded from a trusted server, game client will be verified using a checksum and encryption game data [10].

### 3.3 Middleware

It is a challenge to develop a database engine that is required to run a successful MMOG with millions of players. Therefore, the use of a P2P middleware which enables the development of such complex applications such as MMOG , and efficient entity maintenance and interactions for the highly interactive and visual P2P MMOG application domain.  In most cases, the functionality of the game (such as: replication management, interest management, and update dissemination) is implemented in the game engine having a middleware for programmers. To provide an appropriate programming interface for game designers, while hiding the complexity of the underlying architecture. Thus , many different games can be implemented using the same middleware [10,15].

## 4. Conclusion

In this paper, we first introduced the concept of a MMOG , as well MNOG has introduced many interesting challenges; there has been increasing interest designing P2P for games. whereas P2P architecture improves the scalability of the game, but on the other hand reducing the security, therefore a P2P architecture couples with peer auditing may address many challenges such as vulnerability to cheating. Furthermore, P2P architectures are used in state distribution as well as update dissemination to peers. Many techniques have been proposed to solve these challenges, In addition many of these techniques are dependent on the underlying architecture , but many topics such as cheating remain to be fully addressed.

# References

[1] Armand Bonvicino. *Subscription Numbers.* Available: https://smmobl.wordpress.com/tag/mmog-chart/. Last accessed 23 March 2015.

[2] The Statistics Portal . (2015). *Leading massively multiplayer online (MMO) games worldwide from January to September 2014, by revenue (in million U.S. dollars).* Available: http://www.statista.com/statistics/343075/mmo-games-revenue/. Last accessed 23 March 2015.

[3] Aronson, J., Dead Reckoning: Latency Hiding for Networked Games. 1997, Gamasutra. http://www.gamasutra.com/features/19970919/aronson_01.htm

[4] Greer, J. and Z.B. Simpson, Minimizing Latency in Real-Time Strategy Games, in Game Programming Gems 3, D. Treglia, Editor. 2002, Charles River Media, Inc.: Hingham, Massachusetts. p. 488- 495.

[5] Roger Delano Paul McFarlane. Network software architecture for real-time massively-multiplayer online games. Master's thesis, McGill University, 2005.

[6] Knutsson, B., et al. Peer-to-Peer Support for Massively Multiplayer Games. in INFOCOMM '04. 2004. Hong Kong, China.

[7] Singhal, S. and M. Zyda, Networked Virtual Environments: Design and Implementation. SIGGRAPH Series, ed. S. Spencer. 1999, New York: Addison-Wesley and ACM Press.

[8] CORMAN, A., DOUGLAS, S., SCHACHTE, P., AND TEAGUE, V. 2006. A secure event agreement (sea) protocol for peerto-
peer games. In *Proceedings of the International Conference on Availability, Reliability and Security (ARES'06).* 34–41.

[9] CHEN, B. AND MAHESWARAN, M. 2004a. A cheat controlled protocol for centralized online multiplayer games. In *Proceedings of the International ACM SIGCOMM Workshop on Network and System Support for Games (NETGAMES'04).* ACM Press, New York.

[10] Yahyavi, A. and Kemme, B. 2013. Peer-to-peer architectures for massively multiplayer online games: A survey. ACM Comput. Surv. 46, 1, Article 9 (October 2013), 51 pages. DOI: http://dx.doi.org/10.1145/2522968.2522977

[11] Joshi, R. (2008) "CHEATING AND VIRTUAL CRIMES IN MASSIVELY MULTIPLAYER ONLINE GAMES" Royal University of London. Available online: http://www.ma.rhul.ac.uk/static/techrep/2008/RHUL-MA-2008-06.pdf. /. Last accessed 01 April 2015.

[12] Baughman, N., Levine, B.: Cheat-proof playout for centralized and distributed online games. In: Proc. of IEEE Infocom. (2001)

[13] Bettina Kemme. (2014). *MULTIPLAYER GAMES: THE PERFECT APPLICATION TO EXPLORE SCALABLE AND SECURE DISTRIBUTED DATA MANAGEMENT.* Available: http://db.uwaterloo.ca/db_seminars/notes/Bettina2014.pdf. Last accessed 03 April 2015.

[14] B. B. Yang and H. Garcia-Molina. Designing a super-peer network. In Proceedings of ICDE'03, pages 49–60. IEEE, 2003.

[15] Shanika Karunasekera, Scott Douglas, Egemen Tanin, and Aaron Harwood. *P2P Middleware for Massively Multi-player Online Games.* Available: http://middleware05.objectweb.org/WSProceedings/demos/d6_Karunasekera.pdf. Last accessed 08 April 2015.